# Designing tungsten armoured plasma facing components to pulsed heat loads in magnetic fusion machines

R. Mitteau, M. Diez, M. Firdaouss

**CEA, IRFM, F-13108 Saint-Paul-lez-Durance, France.**

## SUMMARY

A possible design rule for preventing surface damage from thermal transients to solid tungsten armour is proposed and formulated for the plasma facing components (divertor, first wall) of magnetic fusion machines. The rule is based on combined results from laboratory experiments and operating fusion machines, and fundamental engineering principles such as the heat flux factor ($F_{HF}$) and fatigue usage fraction ($F_{UF}$). As an example, the rule would allow $2.10^4$ transient heat loads cycles at a $F_{HF}$ of 10 MJm$^{-2}$s$^{-½}$ before the lifetime is considered exhausted. The formulation of the rule using engineering principles allows combining loads of different magnitudes and various number of cycles. A practical example of the rule usage is provided, illustrating loads combination and how the rule may contribute to the component geometrical design. The proposed rule is only valid for surface loading conditions, hence is not usable for volumetric loading conditions such as runaway electrons. Setting a budget lifetime and a design rule does not preclude actual plasma operation beyond the design lifetime. It is actually normal that experimental devices explore a larger domain than the one defined at the time of the design.

## 1. INTRODUCTION

Tungsten is widely regarded as the best solid armour material for plasma facing components (PFCs) in future magnetic fusion machines like DEMO, CFETR, ARC and others [1-4]. ITER has recently suggested revising its plan to incorporating a tungsten armoured first wall, instead of the current baseline beryllium one. While other advanced PFCs armour concepts such as liquid walls and replenishable walls [5] are proposed, these remain exploratory with low technology readiness levels (TRL) [6]. Solid armour tiles remain most the highest TRL (that is, technologically mature) option for future machines. Tungsten outperforms other solid armour materials due to its high melting point (3410°C), low sputtering yield, and lack of compounds formation with hydrogen isotopes, leading to minimal tritium retention. However, tungsten has drawbacks, such as a limited operational temperature range of 800°C (from the ductile-to-brittle temperature transition of about 400°C to the recrystallization temperature of about 1200°C), a high cooling factor when present as an impurity in the plasma, and adverse interactions with helium under neutron irradiation (blistering, fuzz formation).

Despite its high melting temperature, tungsten surfaces are damaged when used as armour for PFCs in magnetic fusion machines, due to various operating conditions and events. Thermal cycling induces wear through alternating thermal expansion and contraction (Figure 1). This fatigue is further complicated by plasma-wall interaction processes, primarily erosion by sputtering and redeposition, exacerbated by physical, atomic and molecular interactions in the presence of multiple species. Section 2 further discusses tungsten damage in this context and its quantification for engineering design. This work focuses on accounting for tungsten surface damage in designing future fusion machines, particularly regarding component shape design.

Tungsten armoured PFCs, like all engineering components, are designed within practical constraints and engineering rules: component segmentation, material selection, manufacturing techniques, assembly, maintenance needs and design rules. These rules, factors and constraints influence the design, and, ultimately, the geometrical shape, subassembly and arrangement. Design rules aim to control component damage to allow operation for a prescribed duration or number of cycles. Limited damage is acceptable, and actually normal. A component designed for no or negligible damage by end-of-life would be considered "overdesigned".

The design process involves estimating expected loads, assessing progressive damage from these loads, and designing to withstand them. Loads include mechanical (gravity, vibrations), electromagnetic, and thermal loads plus others less relevant here. Thermal loads are especially relevant for PFCs as they pose the main challenge and are a common cause of wear or failure.

Designing PFCs to handle thermal loads is complex due to the design-dependent nature of these loads. The plasma heat load on the PFCs is often the dominant contributor to the thermal load. It comes from the charged particles of the plasma which follow the magnetic field lines intersecting the wall. It is known as the conducto-convective heat load and amount up to 1 GW/m² at the separatrix in future large machines. Even recessed wall components receiving particles from the far scrape off layer are subjected to this conducto-convective load. The conducto-convective heat flux is strongly oriented, following the magnetic field lines with a small incidence angle (<5°, sometimes as small as 1°). Consequently, small changes in component shape significantly alter the incidence angle, causing a strong coupling between shape and load magnitude, making the load design-dependent. The component shape is defined to keep the load within technological limits, such as about 10 MW/m² for steady-state heat loads.

Thermal loads are classified by duration : steady state (permanent) and transient (dynamic). Cooled PFCs are usually designed primarily for steady-state thermal loads, as this is their dominant operation mode. Transient heat loads are considered for the definition of the PFC shapes only to the extent that they do not excessively degrade steady-state performance. The final component surface shape results from balancing the need to limit damage from both steady state and transient loads.

Loads are also categorized by normality : normal or off-normal. Some codes go beyond this over simplistic two-classes categorization, and define up to four load categories [7]. Category-1 is the normal operating case, categories-2 and 3 cover incident and unlikely events, and the category-4 pertains to accidental loading conditions. Specific lifetime budget are associated with each category, with increased damage acceptance for categories -2 and 3, and potential mitigation measures like inspection or repair. Category-4 are usually beyond design basis and are not addressed in design provisions.

The most advanced design code for PFCs is the standard design code for internal components (SDC-IC, [8]). The SDC-IC derives from general nuclear pressure vessels design codes (ASME [9], RCC-MR [10]), and includes additional rules for multilayered components subjected to high energy neutron irradiation at high temperatures. Like all design codes, the SDC-IC conceptual logic stems from a principle of damage prevention, and focuses mainly on avoiding bulk structural damage. It does not explicitly address thermal effects, except for armour to heat sink bonding which relates actually to thermomechanical fatigue. It considers there only the steady-state situation, and is governed by a 'design by experiment' approach. The lack of surface damage criteria makes the SDC-IC incomplete as regards design prescriptions. This papers aims at contributing to curing this missing prescription.

Transient heat loads are unavoidable in magnetic fusion machines. Events such as Loss of plasma control, Vertical Displacement Events (VDEs), Magneto-Hydro Dynamic (MHD) effects, Edge Localized Modes (ELMs), plasma disruptions and Runaway Electron events (REs) among others bring transient heat loads to the PFCs. The associated heat flux densities are orders of magnitude larger than the steady state load, even though the event durations are short and the deposited energy densities might be modest. Many of these transient events fall within the normal operation category (cat-1), although larger disruptions and VDEs are classified as cat-2 and 3. REs are cat-4 events. They produce volumetric damage are not considered for this paper proposed design rule. Transient heat loads are now commonly calculated using various codes and methods [11, 12], providing detailed maps of energy distribution at the PFCs surfaces. Tungsten surface damage is well characterized as a function of loads and number of cycles. All technical tools, data and processes are available for a rational accounting for transient heat loads in PFCs design. A design rule for preventing category-I damage to PFCs armour surfaces from pulsed heat transients is proposed in this paper. Before presenting this rule, further aspects of tungsten surface damage from transient heat loads are explored in the next section.

## 2. TUNGSTEN ARMOUR SURFACE DAMAGE

The progressive surface damage to tungsten under repetitive and increasing transient heat loads is well documented by many authors [13-16]. Damage is qualitatively characterized for increasing heat loads : surface roughening, surface cracks, crack networks, brittle destruction and macro particles loss, local melting, extensive melting up to homogeneous melting, and ultimately melt ejection and boiling (Figure 1).

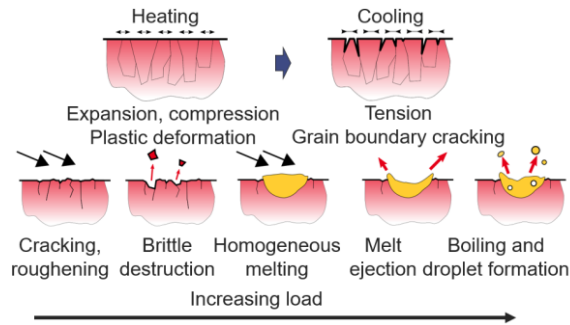

*Figure 1 : Damage to tungsten surface under pulsed heat load*

The key controlling parameter to damage magnitude is the heat flux factor $F_{HF}$. Proposed by J. Linke and others [14, 17], $F_{HF}$ is initially defined for rectangular heat pulses, as the product of heat flux density q (in W/m², or better MW/m²) and the square root of the load duration t : $F_{HF} = q \times \sqrt{t}$. Alternatively, $F_{HF}$ can be expressed as $F_{HF} = e / \sqrt{t}$, where e is the surface energy density (in J/m², or better MJ/m²). Tungsten surface damage primarily depends on $F_{HF}$, even with notable changes of q or t. This is because the surface temperature increase scales proportionally to $F_{HF}$ in the semi-infinite wall approximation, and any finite solid homogenous armour behaves as a semi-infinite wall for short pulses. The physical unit of $F_{HF}$ is $Wm^{-2}s^{½}$ or $Jm^{-2}s^{-½}$ being used in this paper (with the unit prefix "Mega", $MJm^{-2}s^{-½}$). The definition of $F_{HF}$ extends to non-rectangular time waveforms by approximating them to rectangular ones using proxy parameters like time-averaged heat flux or assimilating the pulse duration to the full-width at mid-height of the waveform. The effect of non-rectangular waveforms, though secondary, is described in [18] and $F_{HF}$ remains a valid gauge for surface damage even with non-rectangular waveforms.

Most studies report tungsten surface damage as a function of $F_{HF}$ and the number of load cycles $N_{cycl}$, as shown in Figure 2 which is elaborated from [19] toward a design curve.

Two recent in-machine results are included in Figure 2 : the WEST tungsten divertor surface fatigue point [20-22] and the JET 2013 melt experiment [23], represented as open diamond symbols. The WEST data point reflects the divertor leading edge exposed to all plasma discharges since the WEST's start in 2018. This is $10^4$ plasma discharges, including a significant fraction ending with a disruption (76% as cited in [20], though this rate has reduced). As of 2024, many power discharges still end in disruptions due to various events (so called UFOs, which are eroded material chips from the wall, MHD, medium-Z impurity ingress into the plasma [24]). Publications [20, 21] cite a $F_{HF}$ of 30 MJm$^{-2}$s$^{-½}$ for the WEST divertor leading edge, corroborated by the surface temperature increase measured by the infrared viewing system (IRVS) during thermal quenches. Although the IRVS has a cycle time of 20 ms, it occasionally captures the peak temperature increase during disruption. The divertor temperature increase during disruption often exceeded +100 °C on the PFU front face, at a time where the divertor surface was still rather clean (hence mostly unperturbed from temperature artefacts from dust and codeposited layers observed during later campaigns). A $\Delta$T of 100°C on the divertor PFC bevel surface matches the disruption magnitude number (30 MJm$^{-2}$s$^{-½}$) already written in [21], once recalculated as a $F_{HF}$ on the leading edge. It is estimated here that about 1500 disruptions have occurred with this $F_{HF}$. Most of them occur about at the same place, contributing with other phenomena to the visible alteration of the divertor front surface in WEST (Figure 3).

The JET 2013 melt layer experiment [23] is also shown in Figure 2. The number of events (150 ELMs) is well defined, but the $F_{HF}$ is less certain. Further heat load estimations from co-authors to [23] suggest a value of 40 MJm$^{-2}$s$^{-½}$ [25, 26], albeit with large error bars.

Two calculated $N_{cycl}$ = 1 points are added in Figure 2, from the semi-infinite wall expression : the start of recrystallization event, assimilated to the start of cracking, with a heat flux factor of 17 MJm$^{-2}$s$^{-½}$ (based on reaching the recrystallization temperature, assumed reached for a temperature increase of 1000°C) and the melting heat flux factor of 55 MJm$^{-2}$s$^{-½}$ (based on a temperature increase of 3300°C). These both calculations assume the tungsten thermal conductivity at 130 W/mK, the thermal capacity at 140 J/Kg.K, and the density at 193000 Kg/m$^3$. Both data points are labelled as "calculated points" in figure 2 , and are merely there to indicate the tendency toward low cycle fatigue.

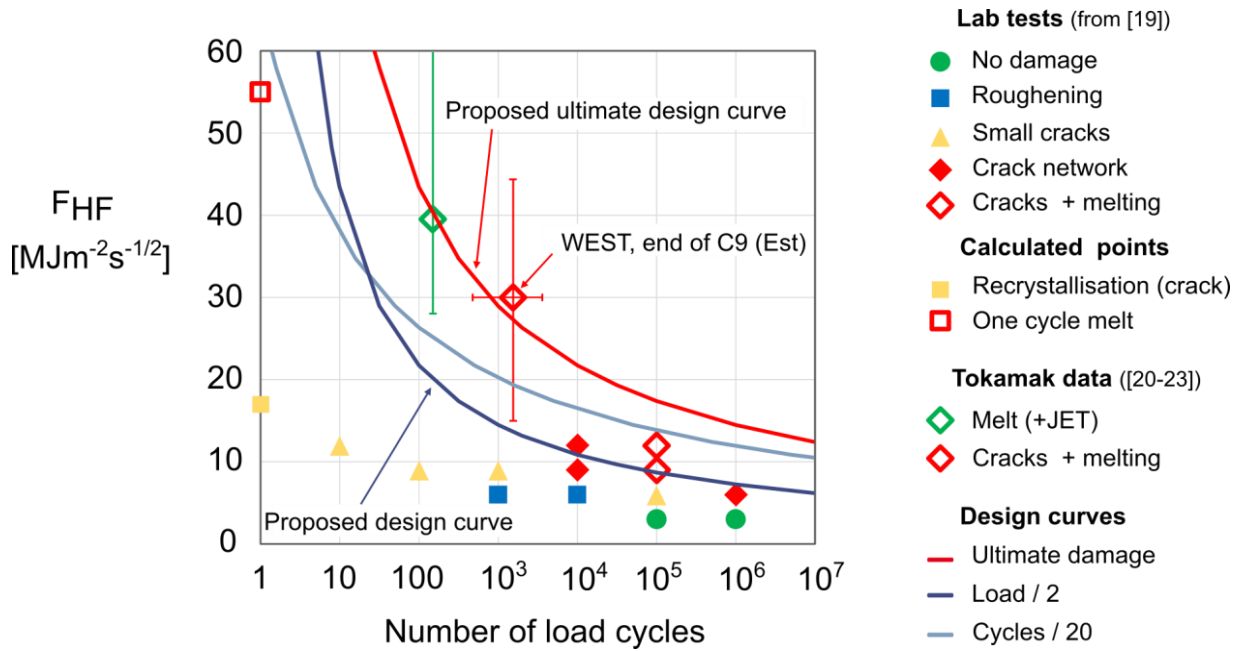

*Figure 2 : Tungsten armour damage expressed as heat flux factor FHF as a function of the load cycle number, along with proposed ultimate damage curve (blue) and design law (Yellow).*

Despite varied conditions and initial temperatures, data in Figure 2 show consistency. In-machine data points from WEST and JET align with laboratory tests, although at much higher damage, indicating good correspondence between lab tests and in-machine observations. Damage curves (crack, melt) expressed as $F_{HF}$ follow a decaying exponential-wise function decreasing for increasing $N_{cyc}$, similar to fatigue damage seen in mechanical fatigue of metallic tensile probes. Thus, deriving fatigue calculation and design rules for surface fatigue damage from mechanical cycling laws is natural.

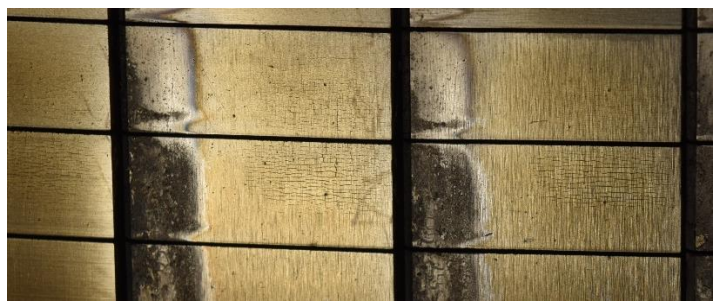

*Figure 3 : Close up view of WEST divertor in 2024 after the C9 experimental campaign, showing the crack network at the surface of some divertor macroblocks. These are 4 monoblocs MB 28 and 29 on units 31 and 32 on the sector 3, and the blocs are 30.5 x 12 mm. The photograph illustrate mainly the front face, loaded with a /20 FHF compared to the edge, barely visible in the picture. Close up view photographs of the leading edge are available in [20] for C4 campaign, and upcoming publications will provide pictures of the leading edge after C9.*

The data in Figure 2 ignore damage dependence on base temperature, reflecting its secondary importance , as also visible in [16]. The secondary importance of base temperature on tungsten surface damage magnitude could be due to damage from higher initial temperature being alleviated by a starting temperature above the ductile to brittle transition temperature, although this is speculative. The modest damage dependency with base temperature supports neglecting the base temperature in the lifetime design curve, and assuming damage accounting solely based on $F_{HF}$ and $N_{cyc}$ for engineering and design purposes. This is also justified by the making of a design rule, which aims at having a simple evaluation method enveloping reasonable experimental findings, while not being excessively constraining. The design law is distinct from a scaling law fitting experimental finding, the latter being expected to be as precise as possible.

## 3. A POSSIBLE RULE TO ACCOUNT FOR TUNGSTEN SURFACE DAMAGE

The method to derive a design rule for cyclic load causing fatigue from experimental data is described in the SDC-IC rule number 2751. First, an ultimate damage curve must be established. This curve expresses the ultimate damage load ($F_{HF}$) as a function of the number of load cycles ($N_{cyc}$). The design curve is then obtained from the ultimate damage curve by taking the smaller of two following curves : either the curve where cycles are divided by 20, or the curve where loads are divided by 2. These discount factors (loads/2 and cycles/20) account for data variability, and ensure that all real components fall within the design curve, rather than serving as a safety margin. Any eventual safety margin would need to be deduced from the design curve.

For the standard tensile probes used in cyclic mechanical fatigue, ultimate damage is straightforward, defined as probe rupture. Defining an ultimate damage for surface fatigue is more complex, as there is no universally recognized criteria for when surface damage renders a fusion machine non-operational. Whether a damaged PFC in a critical location forbids further operation, whether surface brittle destruction causes macro particles loss that penetrates the plasma and practically hinders effective plasma operation, whether degraded plasma operation is pursued despite damage armour, all affect when a component is declared worn out. Current fusion machines (WEST, EAST, ASDEX, JT-60SA, KSTAR) operate with damaged PFCs, and sometimes deliberately induce controlled damage to PFCs armour [21], which is a way of confirming science models. Operation with melted PFCs is however uncertain due to possible run-away situation. The design should strive at maintaining the PFCs in a known state, thus avoiding melting. This paper proposes an ultimate damage curve that balances passing through the data cloud of well molten tiles and ensuring the design curve intersects areas of moderate damage (small cracks, crack network). Engineering components are typically designed to tolerate limited & acceptable damage, provided is remains compatible with effective operation. This approach is common in civil engineering and aerospace, where controlled, acceptable and limited cracking is allowed in functional structural parts.

The most constraining curve is the /2 load curve in Figure 2, leading to the proposed design curve, defined by :

$$F_{HF} = 100/\ln(N_{cycl}) \text{ [in MJm}^{-2}\text{s}^{-½}]$$

For example, a lifetime exhaustion in $10^3$ cycles corresponds to a $F_{HF}$ of 14.5 MJm$^{-2}$s$^{-½}$. For a given load level $F_{HF}$, the unitary usage factor $F_{UF}(F_{HF})$ associated to one single load cycle is $1/N_{cycl}$. As an illustration, $F_{UF}$ ($F_{HF}$ = 14.5 MJm$^{-2}$s$^{-½}$ ) = $1.10^{-3}$, since 1000 cycles at 14.5 MJm$^{-2}$s$^{-½}$ reaches the design lifetime exhaustion. For an arbitrary number of load cycles N, the cumulative fatigue usage factor is N x $F_{UF}$. A cumulative fatigue usage factor less than 1 means that the component is still usable, above than 1 means that the component has exceeded its theoretical lifetime (the component may still be usable, but this is a risky territory). For the above numbers, 100 load cycles at $F_{HF}$ = 14.5 MJm$^{-2}$s$^{-½}$ consume 10% of the available lifetime. The usage factor is dimensionless and additive, allowing loads of various magnitudes to be summed. For these situations of combining loads of varying magnitudes, lifetime is considered exhausted when the cumulative usage factor for all load cycles of different magnitudes reaches 1.

Conventional designing for cyclic damage generally addresses progressive loss of functionality over more than $10^4$ cycles. Fatigue damage leading to loss of functionality in fewer than $10^4$ load cycles (low cycle fatigue) remains an active research domain, with many alternative approaches, including risk and damage management, but no universally accepted practices. This doesn't preclude design activities; it simply indicates an area with significant unknowns and variability, characterised by scattered statistics and potential extreme outliers, reflecting its low TRL.

## 4. ILLUSTRATION OF DESIGN RULE APPLICATION

The usefulness of the proposed design rule for PFC shape design is illustrated using a generic tungsten armoured large tokamak. To focus on methodology and avoid sensitive project-specific details, a representative example case is chosen rather than any given project. This example combines loads from mitigated disruptions and controlled ELMs impacting the same location - the machine ceiling near the secondary upper X-point. While a full realistic exercise shall consider all possible transient events at this location, this simplified loads combination best serves to illustrate the principle.

An hypothetical machine with a major radius R = 6 m, a minor radius a = 2, and a circulating power of approximately 100 MW is assumed. The thermal quench of mitigated disruptions at the vessel ceiling is modelled with $e_{//}$ = 5.2 MJ/m² over 2 ms (values are derived from [27, 28], representative of a large burning fusion machine). An energy decay length of 100 mm for the disruption is assumed, accounting for normal heat flux decay length and energy pattern broadening during the thermal quench, values consistent with the publications referred above. The vessel ceiling also faces ELMs, with a transient thermal load of $e_{//}$ = 1.3 MJ/m² over 0.375 ms, using the same sources. ELMs power pattern assumes a decay length of 50 mm due to less broadening.

The loads are applied to a PFC at the vessel ceiling, assumed to be armoured with tungsten tiles. Assuming a 10% disruptivity over 30000 operation cycles, 3000 disruptions are required. The lifetime is first attributed to disruptions, then the remaining lifetime to the ELMs. Typically, the vessel ceiling is covered with a system of roof shaped components able to handle bidirectional plasma loading (Figure 4), which forbids using simple bevelled components as in a divertor. This shape complicates edge heat load management compared to bevelled PFCs [29]. The PFC has a "front" face which directly faces the plasma, at a small field line incidence angle (between 2 and 15°, assumed here at 5° at the apex), and an "edge" face recessed by a distance $\Delta_{Set\text{-}back}$ (Figure 4, see also [29]).

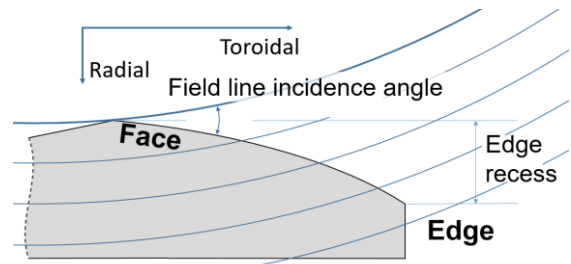

Figure 4 : Sketch of a roof-shaped first wall component for a large tokamak

As shown in [29], full edge shadowing is unattainable for practical design, resulting in some plasma-wetted edges. Even divertors with bevelled shapes allegedly designed for full edge shadowing experience toroidal edge loading, see [30]. The toroidal facing edge receives near perpendicular loading (assumed here to be 90°). Despite the significant recess, making that the edge receives less powerful plasma stream, the edge can experience larger load than the front, due to the larger incidence angle.

The design of shaped components to prescribed magnetic equilibria ensures that wetted edges have a significant recess from the hot magnetic surfaces. The edge recess results first from the radial setback, which is further increased by the mutual shadow in-between the components themselves. Component shape optimisation ensures edge heat loads remain compatible with the budget lifetime [29, 31]. For the present example, $\Delta_{Set\text{-}back}$ is assumed to 50 mm, with shadowing effects tripling the effective recess, to 150 mm. Based on these assumptions, the load and cumulative fatigue usage fraction $F_{UF}$ are calculated for both the front and edge faces for both disruptions and ELMs (Table 1). The number of allowed cycles is also provided. The same edge criterion is used for both face and edge, since surface damage concerns the first tenths to hundreds of µm of the material, and the material behaves essentially like a semi-infinite wall.

|  | $e_{//}$ [MJ/m²] | $\Delta t_{event}$ [ms] | Front face (5° incidence) | | | Edge face (Normal incidence, 150 mm recess) | | |
| --- | --- | --- | --- | --- | --- | --- | --- | --- |
|  |  |  | $F_{HF}$ [MJm$^{-2}$s$^{-½}$] | $N_{cyc}$ | $F_{UF}$ [.] | $F_{HF}$ [MJm$^{-2}$s$^{-½}$] | $N_{cyc}$ | $F_{UF}$ [.] |
| Disruption | 5.2 | 2 | 10.1 | 19274 | 5.2 10$^{-5}$ | 25.9 | 47 | 2.2 10$^{-2}$ |
| ELM | 1.3 | 0.375 | 5.9 | 2.7 10$^7$ | 3.78 10$^{-8}$ | 3.3 | 10$^{13}$ | 1.0 10$^{-13}$ |

*Table 1 : $F_{HF}$, $F_{UF}$ and number of cycles for a disruption and an ELM, both for the component face and edge. $F_{HF}$ comes from $e_{//}$ and $\Delta t_{event}$. $N_{cyc}$ comes from $F_{HF}$ and the design curve, providing $F_{UF}$ by inversion. The numbers are calculated with $\lambda_q = 50$ mm and a disruption broadening of 2.*

In this example, 3000 disruptions result in a cumulative usage fraction of 3000 x 5.2 10$^{-5}$ = 0.156 for the front face. The remaining usage fraction for ELMs on the front face is 1 - 0.156 = 0.844, allowing 0.844 / 3.78 10$^{-8}$ = 22.3 10$^6$ ELMs before the available lifetime is exhausted. For the edge, the lifetime is already exhausted by 47 disruptions, leaving no allowance for ELMs. This could necessitate revising the component shape to increase $\Delta_{Set-back}$, though this would increase front heat loads due to a steeper roof slope, reducing the number of allowed ELMs on the front face. The design rule reveals the necessary compromise between front face and edge damage, enabling rational component shape design.

This simple example shows how transient heat loads can be combined and compared to a traceable budget lifetime. This method can be used alongside those for steady-state heat loads, providing a rational method to optimise PFC design under combined steady-state and transient loading conditions. Note that for a real estimation, other transient heat load events like VDEs, disruptions current quench, and possible unmitigated disruptions should be included, likely resulting in a less favourable outcome and maybe further design considerations.

REs are not included here for two reasons : first, REs dump energy into the tile volume, making the damage-energy relation different from surface heat load, rendering the proposed design rule inapplicable. Second, REs are often considered "beyond design basis" events (cat-4 events), where mitigation solutions like repair & maintenance are preferred over PFC design solutions. Discussion on REs in design processes are beyond this paper's scope.

## 5.  CONCLUSION

A design rule for preventing damage to the tungsten armour from thermal transients is proposed and formulated. A practical example of the rule's application is provided, illustrating its potential feedback on design considerations. The rule is grounded in combined results from laboratory experiments and large scale operating experiments, and fundamental engineering principles.

Formulating a possible design rule raises several key questions that should addressed during the conceptual design of future tungsten armoured fusion devices :

- What thermal transients do affect the main chamber wall – What is their magnitude, location, and frequency? Do transients events combine to cause cumulative damage?
- What kind and extend of surface damage can be tolerated before operation must cease for armour repair or refurbishment ?
- If the effects of heat transients to the main chamber wall are considered a R&D feature and the design proceeds without accounting for them, how should wall maintenance be addressed ? Should it involve robots, hot cells or a spare component strategy ?
- Should the wall surface be explicitly designed to withstand other damage mechanisms, such as erosion and redeposition ?

Answering these questions will have significant implications for the main wall concept, including its thickness, tile/component subdivision, and components shape and role. Additional criteria based on the surface damage caused by sputtering and redeposition could be defined, with relevant controlling parameters to be established.

This rule applies to pure W. Using similar principles, further material-specific design curves (based on dedicated experimental data) would be necessary for next-generation PFC concepts (W alloys, W fiber composites, ceramic materials, etc.) or tungsten coated PFCs.

Finally, setting an budget lifetime and a design rule does not preclude operation beyond these limits. It is expected that experimental devices will explore a broader domain than initially defined at the design stage, attempting extended operation with damaged components as long as other critical requirements, such as nuclear and human safety, are met.

## ACKNOWLEDMENT

The author wishes to thank Sophie Carpentier Chouchana from the ITER organization, who has been inspirational to this work.

## DECLARATION OF GENERATIVE AI AND AI-ASSISTED TECHNOLOGIES IN THE WRITING PROCESS

The authors used ChatGPT-4 during the preparation of this work in order to improve the English style. After using this tool/service, the authors reviewed and edited the content as needed and take full responsibility for the content of the publication.